\documentclass[sigconf,letterpaper,nonacm]{acmart}

\usepackage{algorithm}
\usepackage{algorithmic}
\usepackage{tabularx}
\usepackage{multirow}
\usepackage{color}
\usepackage{graphicx}
\usepackage{xcolor}
\usepackage{colortbl}
\usepackage{tikz}
\usepackage{amsmath}
\usepackage{subfigure}
\usepackage{booktabs}
\usepackage{nameref}
\usepackage{tcolorbox}
\usepackage{adjustbox}

\newcommand{\rot}[1]{\rotatebox[origin=c]{90}{#1}}

\begin{document}

\title{Phishing the Phishers with SpecularNet: Hierarchical Graph Autoencoding for Reference-Free Web Phishing Detection}

\author{Tailai Song}
\affiliation{
 \institution{Politecnico di Torino}
 \city{Turin}
 \country{Italy}}
\email{tailai.song@polito.it}

\author{Pedro Casas}
\affiliation{
  \institution{AIT Austrian Institute of Technology}
  \city{Vienna}
  \country{Austria}}
\email{pedro.casas@ait.ac.at}

\author{Michela Meo}
\affiliation{
  \institution{Politecnico di Torino}
  \city{Turin}
  \country{Italy}}
\email{michela.meo@polito.it}

\renewcommand{\shortauthors}{Song et al.}

\begin{abstract}
Phishing remains the most pervasive threat to the Web, enabling large-scale credential theft and financial fraud through deceptive webpages. While recent reference-based and generative-AI-driven phishing detectors achieve strong accuracy, their reliance on external knowledge bases, cloud services, and complex multimodal pipelines fundamentally limits practicality, scalability, and reproducibility. In contrast, conventional deep learning approaches often fail to generalize to evolving phishing campaigns.

We introduce SpecularNet, a novel lightweight framework for reference-free web phishing detection that demonstrates how carefully designed compact architectures can rival heavyweight systems. SpecularNet operates solely on a webpage’s domain name and HTML structure, modeling the Document Object Model (DOM) as a tree and leveraging a hierarchical graph autoencoding architecture with directional, level-wise message passing. This design captures higher-order structural invariants of phishing webpages while enabling fast, end-to-end inference on standard CPUs.

Extensive evaluation against 13 state of the art phishing detectors, including leading reference-based systems, shows that SpecularNet achieves competitive detection performance with dramatically lower computational cost. On benchmark datasets, it reaches an F1 score of 93.9\%, trailing the best reference-based method by less than 2 percentage points while reducing inference time from several seconds to approximately 20 milliseconds per webpage. 

Field and robustness evaluations further validate SpecularNet under realistic conditions. In real-world deployments, it identifies over 97\% of phishing websites discovered by reference-based detectors. On a newly collected 2026 open-world dataset of 6,000 previously unseen websites (combining Cloudflare rankings and active phishing sites from an industry feed), SpecularNet maintains strong performance despite being trained on data collected nearly five years earlier, demonstrating robust temporal generalization. Moreover, adversarial experiments based on established frameworks for HTML-level manipulation show resilience to realistic HTML evasion attempts, as SpecularNet’s reliance on DOM structure is inherently harder to manipulate without degrading the visual fidelity of phishing pages.
\end{abstract}




\maketitle

\section{Introduction}

Phishing attacks rank among the most pervasive and damaging cyber threats, exploiting deceptive webpages to exfiltrate sensitive information and causing substantial financial losses for consumers as well as reputational damage to service providers. Despite sustained efforts to mitigate the problem, the threat continues to escalate, with reported business losses reaching approximately \$70 billion in 2024, nearly four times the figure recorded in 2023. These trends highlight the limitations of existing phishing defenses and underscore the urgent need for effective, resilient, and broadly deployable phishing detection technologies.

A wide range of approaches has been proposed to combat web phishing, exhibiting varying levels of effectiveness and inherent limitations~\cite{zieni2023phishing}. Broadly, existing methods fall into heuristic-based and learning-based categories. Heuristic approaches typically rely on predefined rules, signature patterns, or blacklists to identify phishing entities, but struggle to keep pace with evolving attacks due to their dependence on static criteria~\cite{alanezi2021phishing, khonji2013phishing}. Learning-based methods, encompassing machine learning (ML) and deep learning (DL) techniques, extract discriminative features from webpage-related content such as URLs to distinguish between benign and malicious webpages~\cite{do2022deep, catal2022applications}. However, these models are often tightly coupled to their training data, limiting their ability to generalize to emerging phishing campaigns. 

Reference-based approaches have recently emerged as the state of the art in phishing detection due to their strong detection performance~\cite{li2024state}. These methods leverage diverse technologies to extract multidimensional information from webpages and compare it against external references to infer malicious intent~\cite{liu2022inferring, liu2024less}, with recent advances further strengthened by Large Language Models (LLMs)~\cite{li2024knowphish}. By incorporating rich cues such as website screenshots and verified domains, reference-based detectors significantly outperform approaches that rely solely on intrinsic webpage features. However, this performance comes at a cost: multidimensional feature extraction, brand matching, continuous reference maintenance, and complex processing pipelines require sophisticated and expensive infrastructures. As a result, such systems remain impractical for end users and inefficient for large-scale, high-throughput phishing detection.

To bridge the gap between the limited generalization of traditional phishing detectors and the high cost and complexity of reference-based systems, we introduce SpecularNet, a novel end-to-end graph neural network that delivers fast, reference-free detection on standard CPUs, achieving strong performance without relying on external knowledge bases or resource-intensive pipelines.

SpecularNet operates solely on a webpage's domain name and HTML structure, explicitly modeling the Document Object Model (DOM) as a tree and exploiting its hierarchical properties through a lightweight graph autoencoding architecture. By mirroring the tree structure around the root node, SpecularNet integrates principles from autoencoding and graph representation learning, enabling hierarchical, level-wise message passing tailored to tree-structured data. SpecularNet goes beyond applying Graph Neural Networks (GNNs) to DOM trees, introducing (i) a hierarchical autoencoder, (ii) directional, level-wise child-parent/parent-child message passing, (iii) a dual reconstruction/supervised objective, and (iv) a hybrid global/node-level decision mechanism. This creates a structural bottleneck that captures higher-order DOM invariants not addressed by prior DOM-GNN/tree-based models.

We evaluate SpecularNet through an extensive and multi-faceted experimental campaign designed to rigorously assess generalization, robustness, and practical viability. The model is trained on data collected in 2021 and evaluated on three independent datasets spanning multiple years (2023–2026), collection sources, and labeling policies, explicitly capturing temporal drift and real-world distribution shift. We benchmark SpecularNet against 13 state of the art phishing detectors across traditional deep learning, generic GNNs, and leading reference-based systems, providing one of the most comprehensive comparative studies to date.

Beyond controlled benchmarks, we validate SpecularNet under realistic deployment conditions. Two field studies assess detection performance on phishing websites observed in the wild, including a newly collected open-world dataset of 6,000 websites combining recent Cloudflare domain rankings and active phishing feeds. Despite a nearly five-year gap between training and evaluation data, SpecularNet retains strong detection performance, demonstrating robust temporal generalization. We further analyze robustness against adversarial HTML-level manipulations using established evasion frameworks, showing that SpecularNet remains resilient to realistic attack strategies that substantially degrade existing learning-based detectors.

Collectively, these results demonstrate that SpecularNet is not only competitive with heavyweight reference-based systems in terms of detection accuracy, but also significantly more efficient, reproducible, and deployable. This comprehensive evaluation provides strong evidence that carefully designed, compact architectures can achieve robust, reference-free phishing detection at scale, making SpecularNet a practical and effective solution for real-world Web security.

% Finally, to support reproducibility, we release our implementation in an anonymized public repository\footnote{\url{https://anonymous.4open.science/r/SpecularNet}}.

\section{Problem Context \& Definition}

In this section, we provide a brief overview of the background with pertinent literature, outline the key drawbacks of reference-based approaches, and formulate the problem.

\subsection{Background}

Phishing comprises fraudulent activities in which attackers deceive users into disclosing sensitive information by impersonating legitimate services. A prevalent enabler of such attacks is the use of phishing kits~\cite{oest2018inside} -- preassembled, ready-to-deploy templates that facilitate the rapid creation of counterfeit webpages and large-scale campaigns. By exploiting users' familiarity with trusted websites and increasingly incorporating obfuscation, redirection, and anti-detection techniques, these kits lower the barrier to entry for attackers and accelerate the evolution of phishing threats in the wild. Existing phishing detection approaches broadly fall into two categories: heuristic-based and ML/DL-based methods.

Heuristic phishing detection approaches rely on predefined rules or attribute-driven analyses, including keyword and HTML structure inspection~\cite{nguyen2014novel, rao2021heuristic}, blacklist and whitelist matching~\cite{azeez2021adopting, openphish, phishtank}, manually crafted rules based on known phishing patterns~\cite{mourtaji2021hybrid, satheeshkumar2022lightweight}, and hybrid systems that combine multiple techniques~\cite{kausar2014hybrid, elgharbi2024online}. While lightweight and interpretable, these methods depend on static criteria and require frequent manual updates, limiting their ability to adapt to evolving phishing strategies.

Learning-based approaches have gained prominence by automatically extracting discriminative features from URLs or HTML content using ML~\cite{DBLP:journals/fgcs/LiYCYL19, venugopal2021detection, nagy2023phishing, choudhary2023machine} and DL models~\cite{le2018urlnet, ozcan2023hybrid, sahingoz2024dephides, Opara_2020, article} trained on large datasets~\cite{catal2022applications, basit2021comprehensive}. More recently, GNN-based methods have exploited webpage structure by modeling DOM trees~\cite{ouyang2021phishing, lindamulage2023vision, wang2024detecting} or incorporating interconnected entities such as links and certificates~\cite{PhishGNN, kim2022phishing}. Despite these advances, existing learning-based detectors often suffer from limited generalizability due to narrow evaluation settings, simplistic webpage encodings vulnerable to unseen content, and insufficient task-specific architectural design.

\subsection{Reference-based Detection}

Reference-based phishing detection encompasses a class of approaches that infer malicious intent by extracting rich, multidimensional representations from webpages and comparing them against external references, such as brand knowledge bases, verified domains, or online resources. Early work primarily leveraged computer vision techniques to identify brand impersonation by matching visual elements (e.g., logos) between phishing and legitimate websites~\cite{abdelnabi2020visualphishnet, rao2015computer, afroz2011phishzoo, lin2021phishpedia}. Subsequent research has significantly expanded this paradigm by incorporating refined analyses of critical webpage components such as credential-requiring pages~\cite{liu2022inferring}, leveraging auxiliary information retrieved from search engines~\cite{jain2018two, rao2019jail}, and constructing continuously updated reference repositories~\cite{liu2023knowledge, li2024knowphish}. More recent advances further integrate large language models~\cite{liu2024less, lee2024multimodal}, AI agents~\cite{wang2024automated, cao2025phishagent}, and multimodal feature extraction combining textual, visual, and structural cues~\cite{lee2024multimodal, zhang2021multiphish}, leading to state of the art detection performance.

Despite their effectiveness, reference-based approaches incur substantial computational and operational overhead due to complex processing pipelines, external dependencies, and costly infrastructures, limiting their scalability, accessibility, and reproducibility. These constraints hinder practical deployment for end users and large-scale, real-time phishing defense, and complicate fair benchmarking and reproducible research~\cite{li2025phishintel, petrukha2024position}.

\subsection{Problem Statement \& Threat Model}

We formulate a binary classification problem aimed at developing an end-to-end model that overcomes the limitations of existing phishing detection approaches, addressing the shortcomings of conventional methods while rivaling the effectiveness of reference-based systems without their associated overhead. Formally, given a webpage, we design a deep learning model that relies solely on its domain name and HTML structure to predict whether it is benign or phishing. Consequently, our approach is intended as a lightweight phishing detector that can be readily integrated into existing frameworks for both end users (e.g., browsers or email clients) and service providers (e.g., detector ensembles or defense pipelines).

In terms of threat model, we consider large-scale web phishing attacks in which adversaries deploy counterfeit webpages using reusable phishing kits or automatically generated templates. Attackers are assumed to control the domain name and HTML structure of the webpage, and may apply common obfuscation or evasion techniques at the HTML level (e.g., tag reordering, insertion of redundant nodes, or superficial structural perturbations).

We assume no access to external references, brand knowledge bases, screenshots, network context, or historical reputation signals at inference time. Detection is performed solely on the webpage's domain name and DOM structure, under a resource-constrained setting suitable for CPU-only execution.

We do not consider highly targeted spear-phishing attacks requiring personalized content, user interaction history, or multi-step behavioral analysis, nor do we assume a fully adaptive white-box adversary capable of arbitrarily redesigning webpage structure without degrading visual fidelity or usability. Within this scope, SpecularNet aims to provide robust, reference-free phishing detection that generalizes across time and evolving campaigns.

\begin{figure}[t!]
\centering
\includegraphics[scale=0.43]{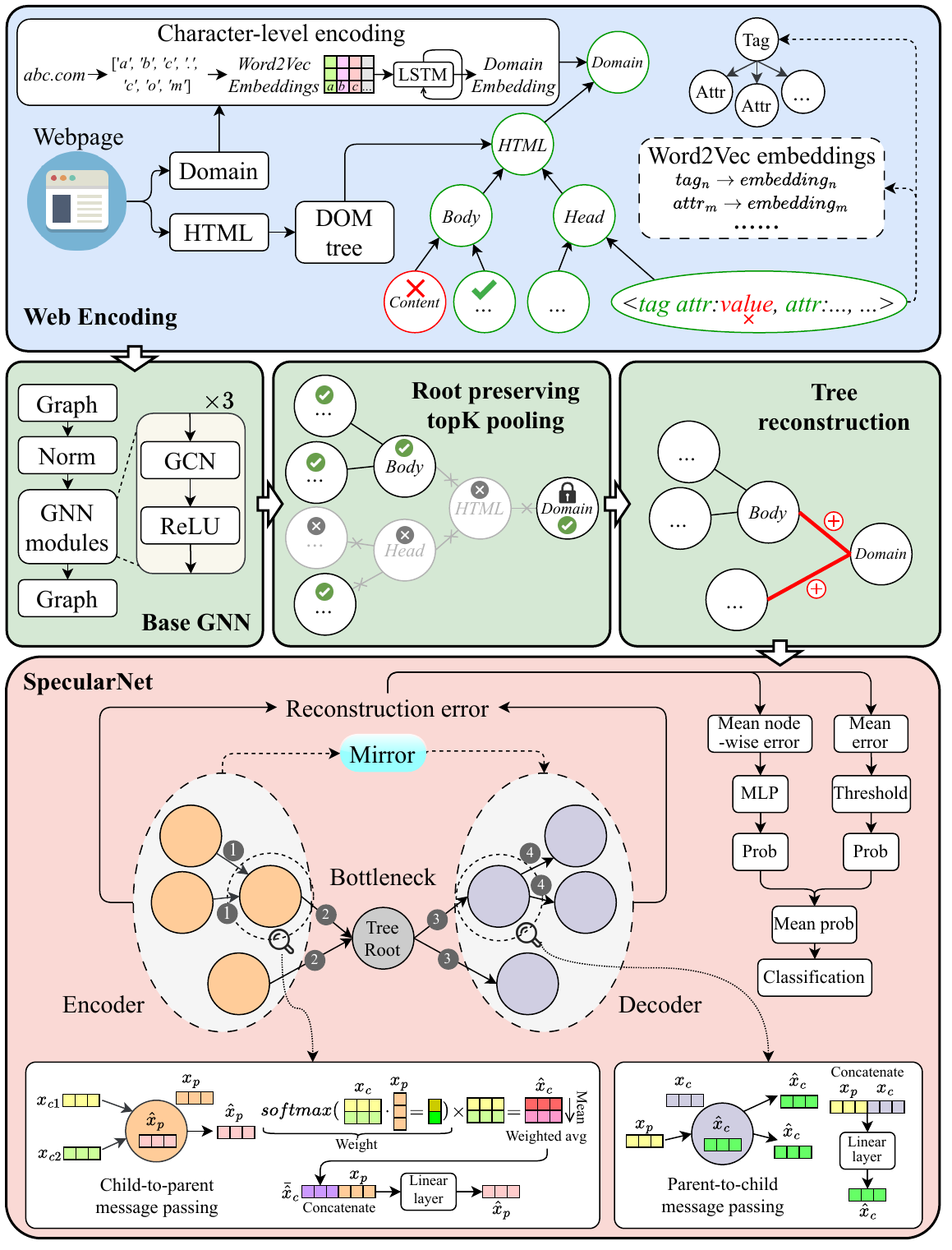}
\vspace{-1.5em}
\caption{Overview of the SpecularNet framework.}
\label{fig:arch}
\end{figure}

\section{The SpecularNet Framework}

Figure~\ref{fig:arch} provides an overview of SpecularNet. We first motivate the design choices and highlight the key novelties, and then describe the architecture in detail.

\subsection{Rationale \& Relation to Prior Work}\label{sec:relation}

SpecularNet's design is guided by four key observations. First, phishing webpages are commonly generated using phishing kits that introduce distinctive domain patterns and HTML semantics seldom present in benign websites. Second, the HTML Document Object Model (DOM) inherently forms a tree-structured graph, making it particularly well suited for graph neural networks (GNNs) to capture structural dependencies and discriminate between benign and malicious webpages. Third, we deliberately avoid encoding arbitrary textual content, which is often noisy and highly variable, to improve generalization to previously unseen phishing patterns. Finally, instead of relying solely on conventional supervised learning, we incorporate elements of quasi-unsupervised and ensemble learning to mitigate overfitting.

In relation to prior work, although tree structures are well studied, they remain underexplored as first-class inputs to deep learning models in phishing detection. Several studies~\cite{qiao2020tree, wang2021tree, talak2021neural} employ tree representations by decomposing general graphs into subtrees, but their objective remains graph-level representation learning rather than explicitly modeling trees as the primary data structure. Other tree-oriented neural models~\cite{tai2015improved, wang2023learning, cheng2018treenet, yu2022tnn, hu2023synctree} focus on different tasks (e.g., node classification) or domains (e.g., circuit design), and are therefore not directly applicable to webpage analysis.

SpecularNet differs fundamentally from these lines of work by treating the DOM as a \emph{dedicated tree} and designing a task-specific architecture that explicitly exploits its hierarchical structure. While superficially related concepts such as hierarchical pooling~\cite{zhang2019hierarchical} or hierarchical message passing~\cite{zhang2019dvaevariationalautoencoderdirected, noncomputable2023molecules, qiao2020tree} have been explored in prior studies, their goals and mechanisms differ substantially. Hierarchical pooling aims to progressively coarsen graphs to obtain compact representations, whereas SpecularNet preserves the full tree structure and leverages hierarchy to enable directional, level-wise information flow. Similarly, hierarchical message passing in existing models is typically introduced as a design heuristic, while in SpecularNet it arises naturally from the tree structure and the autoencoding formulation.

Although modeling the DOM as a graph is common practice, SpecularNet advances the state of the art in three key ways: \textit{(i)} it explicitly models the DOM as a tree rather than a generic graph, \textit{(ii)} it addresses known deficiencies in existing webpage encodings that hinder generalization, and \textit{(iii)} it adapts the tree structure itself to facilitate more effective encoding and message passing. Importantly, the novelty of SpecularNet extends beyond DOM modeling alone, introducing a principled architectural framework tailored to phishing detection.

Finally, SpecularNet should not be confused with graph autoencoders such as the Graph Variational AutoEncoder (GVAE)~\cite{kipf2016variational} or prior tree-based autoencoders~\cite{jin2018junction, irsoy2017unsupervised, alvarez2017tree}. While these models focus on learning stochastic latent distributions for graph generation, SpecularNet adopts a deterministic autoencoding strategy that exploits *structural duplication* to reconstruct or distort tree-structured inputs for discrimination. To further substantiate this distinction, we include an advanced GVAE variant, PIGVAE~\cite{winter2021permutation}, as a strong representation-learning baseline in Section~\ref{sec:setup}. Overall, SpecularNet is the first model to explicitly couple hierarchical tree structure with graph autoencoding to enable task-driven, directional message passing for phishing detection.

\subsection{Webpage Encoding}

SpecularNet relies exclusively on a webpage's domain name and HTML structure for phishing detection. Prior work has shown that URLs can encode discriminative signals between benign and malicious webpages~\cite{li2019stacking, aung2019survey}. However, many existing approaches implicitly exploit URL length as a dominant feature by contrasting typically long phishing URLs with concise homepage URLs of legitimate sites, yielding inflated and brittle performance. To avoid this bias, we focus on domain names, which are largely length-insensitive due to standardized termination (e.g., \texttt{.com}), and aim to learn latent patterns without manual feature engineering.

We adopt a character-level encoding for domains: each domain is represented as a sequence of characters drawn from a fixed ASCII vocabulary (approximately 60 tokens). Characters are embedded and composed into a domain-level representation using a single-layer LSTM~\cite{hochreiter1997long}. This design eliminates out-of-vocabulary issues inherent to word-level embeddings and ensures robust generalization to unseen domains. Unlike approaches that rely on pretrained word embeddings, our domain encoding operates purely at the syntactic level, reflecting the constrained structure of domain names.

For HTML, SpecularNet exploits the inherent structure of the DOM tree. Each non-leaf node corresponds to an HTML tag accompanied by one or more attribute-value pairs, while leaf nodes contain webpage content. We deliberately encode only tags and attributes, discarding attribute values and textual content, for two reasons. First, values and contents form an unbounded and highly variable vocabulary, comparable to natural language, making tokenization impractical and severely limiting generalization given the finite training corpus. In contrast, tags and attributes are drawn from a largely fixed and standardized set defined by the HTML specification; although custom elements may appear, they are rare and have negligible impact. Second, phishing webpages are designed to visually mimic legitimate sites, rendering user-facing content largely uninformative or even misleading for detection.

To simplify representation, each DOM node is decomposed into a small subtree in which the tag serves as the parent and its attributes are modeled as child nodes, avoiding the complexity of jointly encoding multiple entities within a single node. Tags and attributes are mapped to embeddings using Word2Vec, trained only on documented HTML-standard elements observed in the training set, while unseen tags or attributes are mapped to dedicated unknown tokens. The ordered sequence of a tag and its attributes forms a sentence for embedding generation. Finally, the domain embedding is attached to the DOM root, yielding a unified tree representation. Although Word2Vec is a relatively lightweight embedding method, it is sufficient in this context, as SpecularNet relies on structural modeling rather than semantic understanding to capture discriminative patterns.

\subsection{A Tailored Graph Neural Network}

The resulting representation of a webpage is a tree whose nodes correspond to either tags, attributes, or the domain embedding, and whose edges encode parent–child relationships without edge features. While this structure naturally lends itself to graph neural networks, generic GNN architectures are not well aligned with tree-structured data, nor with the specific requirements of phishing detection. We therefore design a tailored, task-specific model that builds upon established components while explicitly respecting the tree topology.

First, node features are standardized via layer normalization and processed by a stack of three Graph Convolutional Network (GCN) layers~\cite{kipf2016semi}, each followed by a LeakyReLU activation~\cite{xu2020reluplex}, to refine node representations and capture both local and global structural dependencies. Second, to remove uninformative nodes and improve computational efficiency for downstream processing, we apply top-$k$ pooling~\cite{knyazev2019understanding} to select a compact subset of nodes, while always retaining the root node to preserve the global structure\footnote{If multiple nodes remain at the top level after pooling, selecting an alternative root or modifying the bottleneck would introduce additional complexity and is left for future work.}. Finally, because top-$k$ pooling may disrupt tree connectivity, we explicitly restore the original parent-child relationships by tracing each retained node's ancestral links.

\subsection{The SpecularNet Architecture}

SpecularNet adopts an AutoEncoder(AE)-inspired architecture tailored to tree-structured data. For clarity, the terms \emph{encoder} and \emph{decoder} refer exclusively to the two complementary halves of our AE-style design, and not to generic graph neural networks. In the encoder stage, the DOM tree is progressively aggregated toward the root, whose position naturally serves as a bottleneck for learning compact structural representations. For decoding, we construct a \emph{specular} counterpart of the original tree by mirroring its non-root nodes and attaching this reflected structure to the root, enabling controlled reconstruction of the input topology.

Unlike conventional autoencoders, where compression and reconstruction are achieved through successive dimension-changing layers, SpecularNet integrates representation learning directly into the graph structure itself. Consequently, encoding and decoding are not realized through standard feed-forward transformations. Instead, we introduce hierarchical, directional message passing, propagating information bottom-up during the encoding phase and top-down during decoding. This design aligns naturally with the tree topology and enables effective structural reconstruction.

\subsubsection{Encoder}

The node features and connectivity used in the encoder are inherited from the preceding stages of the framework. The encoder updates node representations through a linear transformation and a structured message aggregation process. Unlike conventional GNNs that propagate information symmetrically among neighboring nodes, SpecularNet employs \emph{directional message passing}, in which messages flow exclusively from child nodes to their parent.

Let the feature vector of a parent node \(P\) be \(x_p \in \mathbb{R}^F\), where \(F\) denotes the feature dimension, and let \(x_c \in \mathbb{R}^{N \times F}\) denote the feature matrix of its \(N\) child nodes \(C_i, i \in [1,N]\). We first compute the relative importance of each incoming message by applying a Softmax function to the dot product between child features and the parent feature:
\begin{align}
    w_c = Softmax(x_c \cdot x_p^T)
\end{align}

The resulting weights are then applied to the child features, and the weighted messages are aggregated via averaging to obtain a condensed representation:
\begin{align}
    \bar{\hat{x}}_c = \frac{1}{N} \sum_{i=1}^{N} \hat{x}_c, \text{with} \ \hat{x}_c = w_c \times x_c
\end{align}

Finally, the parent node feature is updated by concatenating the aggregated message \(\bar{\hat{x}}_c\) with the original parent feature \(x_p\), followed by a linear transformation:
\begin{align}
    \hat{x}_p = \sigma(W_{encoder} \cdot \bar{\hat{x}}_c||x_p + b_{encoder})
\end{align}
where \(W_{encoder}\) and \(b_{encoder}\) denote the learnable weights and bias of the encoder’s linear layer, and \(\sigma\) is the LeakyReLU activation function. For leaf nodes without children, a single zero vector is used as the incoming message. The same update procedure is also applied to the root node.

\subsubsection{Decoder}

The decoder mirrors the encoder in structure but operates in the reverse direction. Message passing proceeds from parent nodes to their children, following the inherent hierarchy of the tree. The decoder reuses the original tree topology, while node features are initialized using their original embeddings prior to encoding. This design enables a controlled reconstruction of node representations based on information propagated from higher levels of the tree.

Because each child node has exactly one parent in a tree, each node receives a single incoming message during decoding. Consequently, message aggregation is not required, and node updates can be performed directly. Let \(x_p\) denote the feature vector of a parent node and \(x_c\) the feature vector of its child node. The updated child representation is computed as:
\begin{align}
    \hat{x}_c = \sigma(W_{decoder} \cdot x_p||x_c + b_{decoder})
\end{align}
where \(W_{decoder}\) and \(b_{decoder}\) are the learnable weights and bias of the decoder’s linear layer, respectively, and \(\sigma\) denotes the LeakyReLU activation function.

During decoding, node features act as latent states that are progressively refined as information flows from ancestors to descendants. Message passing therefore follows a hierarchical, level-wise order, enabling structured reconstruction of the tree representation. The root node serves as the central redistribution point, and node updates are performed sequentially across levels, ensuring that each node is processed only after its parent has been updated. This asynchronous update scheme contrasts with conventional GNNs, which typically update all nodes simultaneously. 

\subsubsection{Ensemble Classification}

To go beyond a standard AE decision rule, we couple reconstruction-based scoring with a learned discriminator on reconstruction errors. Specifically, we map reconstruction errors to two probabilities: one derived from a calibrated threshold on the scalar reconstruction error, and another produced by an MLP operating on an aggregated error vector. The final prediction combines both confidences, yielding a more stable classifier than either component alone.

Let \(X \in \mathbb{R}^{N \times F}\) and \(\hat{X} \in \mathbb{R}^{N \times F}\) denote the node-feature matrices of the encoder-side tree and its decoder-side reconstruction, respectively. After hierarchical message passing, we compute the element-wise squared reconstruction error:
\[
E = (X - \hat{X})^2
\]

\noindent\textbf{Reconstruction-based probability:} 
we first compute the mean squared reconstruction error for an input tree:
\begin{align}
    \varepsilon = \frac{1}{NF}\sum_{i=1}^N\sum_{j=1}^F E_{i,j}
\end{align}
Conventional anomaly-detection AEs trained only on benign data are known to fail under distribution shift and memorization effects~\cite{gong2019memorizing, park2021wrong}. Following~\cite{yamanaka2019autoencoding}, we instead exploit class labels \(y\) during training and define:
\begin{align}
    \mathcal{L}_1 = y \cdot \varepsilon - (1-y) \cdot log(1-e^{-\varepsilon})
\end{align}
This objective encourages accurate reconstruction (low \(\varepsilon\)) for benign samples and distorted reconstruction (high \(\varepsilon\)) for phishing samples, resulting in a quasi-unsupervised reconstruction pipeline guided by supervision. After training, we use the validation set to determine a reconstruction threshold \(\tau\). Rather than applying a hard threshold, we convert the distance to \(\tau\) into a calibrated confidence via:
\begin{align}
    prob_1 = \frac{1}{1+e^{\beta \cdot (\varepsilon-\tau)}}
\end{align}
which corresponds to a sigmoid centered at \(\varepsilon=\tau\), with steepness controlled by \(\beta\). \textbf{MLP-based probability:}
in addition, we learn a discriminative mapping from reconstruction errors to class likelihood. We first aggregate errors across the feature dimension to obtain a per-node error summary, and then average over nodes to form an error vector:
\begin{align}
    \delta = \frac{1}{N}\sum_{i=1}^N E_{i}
\end{align}
The vector \(\delta\) is then passed to a two-layer MLP to produce a probability:
\begin{align}
    prob_2 = Sigmoid(W_2 \cdot (\sigma(W_1 \cdot \delta + b_1))+b_2)
\end{align}
where \(W_{1,2}\) and \(b_{1,2}\) are learnable weights and biases, and \(\sigma\) denotes LeakyReLU. We train this branch with binary cross entropy:
\begin{align}
\mathcal{L}_2=\ell_{\text{BCE}}(\hat{y}, y) \text{ with }\hat{y}=\lfloor prob_2 \rceil
\end{align}

\noindent\textbf{Ensemble prediction and joint optimization:} the final decision combines both probabilities:
\begin{align}
    \hat{y}=\begin{cases}0, \ \text{if } (prob_1+prob_2)/2 \leq 0.5 \\1, \ \text{otherwise}\end{cases}
\end{align}
To balance the two training signals, we do not simply sum \(\mathcal{L}_1\) and \(\mathcal{L}_2\). Instead, we adopt learnable loss weights~\cite{kendall2018multi} \((w_1,w_2)\) and optimize:
\begin{align}
    \mathcal{L}_{final}=\mathcal{L}_1 \cdot e^{-w_1} + w_1 + \mathcal{L}_2 \cdot e^{-w_2} + w_2
\end{align}
This formulation yields a multitask objective in which the two branches are complementary: \(\mathcal{L}_1\) shapes the reconstruction space by separating benign and phishing samples via their errors, while \(\mathcal{L}_2\) learns an explicit decision boundary from the resulting error patterns. Because both branches are optimized toward the same direction---low reconstruction error for benign samples and high reconstruction error for phishing ones---their combined confidence provides a robust and consistent classifier.

\begin{table}
\footnotesize
\centering
\renewcommand{\arraystretch}{0.8}
\caption{Details on SpecularNet hyper-parameters.}
\vspace{-1em}
\scalebox{0.97}{
\begin{tabular}{p{4.5cm}|p{1.3cm}<{\centering}}
\toprule
\textbf{Parameter} & \textbf{Value} \\
\midrule  
Feature dimension, $F$ & 32 \\
Num layers of GCN & 3 \\
Size of hidden channel in GCN & 64 \\
Num layers of LSTM & 1 \\
Size of hidden channel in LSTM & 32 \\
Ratio for topK pooling & 20\% \\
$\beta$ converting threshold to probability & 1.0 \& 10.0 \\
Num neurons of linear layer in AE & 32 \\
Num neurons of classification MLP & 16, 1 \\
Activation function & LeakyReLU \\
Optimizer & Adam  \\
Batch size & 8 \\
Initial learning rate & $1\times 10^{-3}$ \\
\bottomrule
\end{tabular}}
\label{tab:hyper}
\end{table}

\section{Experimental Evaluation}

We first describe the implementation details of SpecularNet. We then present the experimental setup and report results covering overall detection performance, ablation analysis, effectiveness against real-world phishing in the wild, and robustness to adversarial attacks

\subsection{Implementation Details}

SpecularNet is implemented in PyTorch, with modules constructed from scratch. Table~\ref{tab:hyper} enumerates each parameter. Note that: \textit{1)} We adopt a cosine annealing schedule with warm restarts for the learning rate. \textit{2)} $\beta$ is determined based on the validation set and differs between models with ($\beta=1.0$) and without ($\beta=10.0$) domain name features. Since the model is retrained in the absence of domain information for test set 2 (explained in the following section), it is reasonable for a different optimal $\beta$ to emerge. Nonetheless, the effect of $\beta$ is minor; for instance, keeping it to 1.0 results in less than a 1\% drop in F1-score. \textit{3)} Hyperparameter tuning follows a standard, empirical pipeline, in which random search is employed to identify the optimal configuration based on validation set performance.

SpecularNet is designed for efficient execution on standard CPUs and does not require GPU acceleration. In practice, CPU-based inference is faster than GPU execution, yielding a substantial practical advantage for large-scale deployment, retraining, and reproducible evaluation. This behavior stems from the model's computational profile: its dominant operations (tree reconstruction and asynchronous, level-wise message passing over sparse DOM graphs) are inherently sequential and memory-bound, and therefore do not benefit from GPU parallelism. SpecularNet is trained and evaluated on an Intel Xeon Gold 6140 server-tier CPU.

\subsection{Datasets and Baselines}\label{sec:setup}

To ensure a fair and direct comparison against reference-based phishing detectors, we take three datasets published by prior work on reference-based approaches with benign/phishing webpages collected from diverse sources over different periods of time:\\
\noindent $\bullet$ The first dataset~\cite{lin2021phishpedia} (training set), collected in 2021, consists of 29,951 benign websites sourced from the top-ranked Alexa list~\cite{alexa} and 29,496 phishing webpages obtained via OpenPhish Premium Service~\cite{openphish}. We refer to it as the \emph{phishpedia-dataset}.

\noindent $\bullet$ The second dataset (test set 1) is curated in 2023 and adopted by two recent reference-based methods~\cite{li2024knowphish, cao2025phishagent}. It contains a balanced collection of 5,000 benign samples from the Tranco list~\cite{pochat2018tranco} and 5,000 malicious samples from PhishTank~\cite{phishtank}. We refer to it as the \emph{knowphish-dataset}.

\noindent $\bullet$ The third dataset (test set 2) is released by~\cite{liu2023knowledge}, collected during a similar period from 2023 to 2024, and used by~\cite{liu2024less}. It comprises 6,075 webpages for both classes, with benign samples crawled from Alexa lower-ranked 5,000 to 15,000 websites~\cite{alexa} and phishing samples created by various validated phishing kits. We refer to it as the \emph{phishllm-dataset}.

Rather than adopting a conventional random train-test split, we train SpecularNet exclusively on the oldest dataset -- \emph{phishpedia-dataset} (2021), reserving 20\% of it for validation, hyperparameter tuning, and selection of the reconstruction threshold $\tau$. The remaining two datasets \emph{knowphish-dataset} and \emph{phishllm-dataset} (2023-2024) are used solely for evaluation purposes. Although the datasets originate from distinct sources and time periods, we additionally verify that no overlaps exist across them, eliminating any risk of data leakage. This protocol rigorously evaluates out-of-distribution generalization under temporal, source, and labeling-policy shifts, using independent datasets collected in 2023-2024 while training in 2021 data. Finally, because the \emph{phishllm-dataset} partially lacks domain name information, we retrain SpecularNet using the same configuration while omitting domain-dependent components.

To demonstrate SpecularNet's performance relative to state of the art detectors across multiple model families, we take a comprehensive benchmark of 13 state of th -art systems, extending beyond mere comparisons with reference-based methods, answering the following three questions:

\noindent $\bullet$ \textit{Do traditional DL-based phishing detectors truly underperform?} We choose four top-performing, applicable DL models explicitly designed for phishing detection: HTMLPhish~\cite{opara2020htmlphish}, Web2Vec~\cite{feng2020web2vec}, HTML-GNN~\cite{ouyang2021phishing}, and PhishDet~\cite{ariyadasa2022combining}. They leverage various techniques such as RNN and GNN and multidimensional features including URLs and HTML contents. 

\noindent $\bullet$ \textit{Can we apply existing general GNNs instead of developing a novel, specialized one?} We meticulously identify three state of the art GNNs, each exhibiting distinct characteristics relevant to our problem: DAG~\cite{luo2023transformers}, a Transformer-based GNN tailored to directed acyclic graphs (as tree structure belongs to such graphs), GNN$^+$~\cite{luo2025unlocking}, a recently-released framework that integrates a suite of best practices to deliver superior performance, and PIGVAE~\cite{winter2021permutation}, the permutation-invariant graph variational autoencoder (cf. Section~\ref{sec:relation}).

\noindent $\bullet$ \textit{Can SpecularNet compete with sophisticated reference-based approaches?} We consider three state of the art frameworks -- KnowPhish detector (KPD)~\cite{li2024knowphish}, PhishAgent~\cite{cao2025phishagent}, and PhishLLM~\cite{liu2024less}, which report the detection performance (on test sets we adopt) for themselves and for a group of other established reference-based methods: PhishPedia~\cite{lin2021phishpedia}, PhishIntention~\cite{liu2022inferring}, GEPAgent~\cite{wang2024automated}, and ChatPhish~\cite{koide2023detecting}. In the evaluations we present next, multiple results for each of these models indicate detectors backed by different knowledge bases.

\begin{table}[t!]
\centering
\caption{Performance on \emph{knowphish-dataset} -- test set 1.}
\vspace{-1em}
\label{tab:perf_1}
\renewcommand{\arraystretch}{1.1}
\begin{adjustbox}{width=0.9\columnwidth}
\begin{tabular}{llccccc}
\toprule
 & \textbf{Phishing detector} & \textbf{ACC} & \textbf{F1} & \textbf{Precision} & \textbf{Recall} & \textbf{Time} \\
\midrule
\multirow{4}{*}{\rot{Traditional}} 
 & HTMLPhish~\cite{opara2020htmlphish}    & 52.01 & 39.80 & 60.33 & 51.96 & $\ll 0.1$ s \\
 & Web2Vecc~\cite{feng2020web2vec}        & 57.21 & 48.45 & 72.24 & 57.24 & $\approx 0.1$ s \\
 & HTML-GNN~\cite{ouyang2021phishing}     & 84.74 & 84.69 & 84.80 & 84.69 & $< 0.1$ s \\
 & PhishDet~\cite{ariyadasa2022combining} & 78.64 & 77.78 & 83.22 & 78.63 & $\ll 0.1$ s \\
\midrule
\multirow{3}{*}{\rot{GNN}}
 & DAG~\cite{luo2023transformers}         & 85.50 & 85.48 & 86.11 & 85.51 & $\ll 0.1$ s \\
 & GNN$^{+}$~\cite{luo2025unlocking}      & 88.39 & 88.42 & 88.66 & 88.42 & $\ll 0.1$ s \\
 & PIGVAE~\cite{winter2021permutation}    & 84.07 & 83.97 & 84.85 & 84.06 & $\approx 0.4$ s \\
\midrule
\multirow{12}{*}{\rot{Reference-based}}
 &                                         & 69.91 & 57.17 & 99.16 & 40.16 & $\approx 0.3$ s \\
 & Phishpedia~\cite{lin2021phishpedia}     & 66.40 & 52.52 & 89.50 & 37.16 & $> 10$ s \\
 &                                         & 85.79 & 83.67 & 98.27 & 78.00 & $\approx 0.2$ s \\\cmidrule(lr){2-7}
 &                                         & 66.62 & 49.96 & 99.76 & 33.32 & $\approx 0.3$ s \\
 & PhishIntention~\cite{liu2022inferring}  & 62.51 & 41.16 & 95.62 & 26.22 & $> 10$ s \\
 &                                         & 77.84 & 71.60 & 99.67 & 55.84 & $\approx 0.3$ s \\\cmidrule(lr){2-7}
 & \multirow{2}{*}[-0ex]{KnowPhish~\cite{li2024knowphish}} 
                                           & 76.10 & 69.71 & 95.16 & 55.00 & $> 10$ s \\
 &                                         & 92.49 & 92.05 & 97.84 & 86.90 & $\approx 2$ s \\\cmidrule(lr){2-7}
 & GEPAagent~\cite{wang2024automated}      & 90.93 & 90.85 & 92.12 & 89.66 & $> 10$ s \\
 & ChatPhish~\cite{koide2023detecting}     & 92.96 & 92.90 & 93.08 & 92.80 & $\approx 7$ s \\
 & \textcolor{red}{PhishAgent~\cite{cao2025phishagent}} & \textcolor{red}{95.36} & \textcolor{red}{95.37} & \textcolor{red}{95.11} & \textcolor{red}{95.64} & \textcolor{red}{\underline{$\mathbf{\approx}$ \textbf{2 s}}} \\
%\midrule
%& \textcolor{blue}{PHISHGRAPH} & \textcolor{blue}{94.29} & \textcolor{blue}{94.22} & \textcolor{blue}{95.33} & \textcolor{blue}{93.14} & \textcolor{blue}{\underline{$\mathbf{\approx}$ \textbf{0.1 s}}} \\
\midrule
& \textcolor{blue}{SpecularNet} & \textcolor{blue}{93.92} & \textcolor{blue}{93.92} & \textcolor{blue}{94.02} & \textcolor{blue}{93.91} & \textcolor{blue}{\underline{\textbf{20 ms}}} \\
\bottomrule
\end{tabular}
\end{adjustbox}
\end{table}

Furthermore, for both categories of traditional DL-based detectors and general GNNs, we adopt their default, suggested configurations, training all models on a single GPU of NVIDIA V100 32GB and tuning their performance using the validation set. Notably, certain existing acclaimed graph-based phishing detection methods~\cite{PhishGNN, kim2022phishing, tan2020graph} are not applicable to our scenarios due to their reliance on specific requirements that are not available in our dataset\footnote{For instance, PhishGNN~\cite{PhishGNN} does not operate directly on webpages but relies on peripheral entities such as outbound links, which fall outside the scope of our work.}. On the other hand, due to the complexities involved, implementing/reproducing reference-based approaches is barely feasible. Since the evaluations in their original publications~\cite{cao2025phishagent, liu2024less, li2024knowphish} were conducted on almost identical datasets, we directly quote their reported results, applying minor adjustments to account for occasional incompleteness in datasets or metrics\footnote{For instance, the work in~\cite{liu2024less} reports only the precision and recall for phishing samples; however, since the quantities of both classes are provided, we can reliably infer the overall performance metrics needed herein.}.

\begin{table}[t!]
\scriptsize
\centering
\caption{Performance on \emph{phishllm-dataset} -- test set 2.}
\vspace{-1.5em}
\renewcommand{\arraystretch}{1.1}
\begin{adjustbox}{width=0.9\columnwidth}
\begin{tabular}{llccccc}
\toprule
\multicolumn{2}{c}{\textbf{Phishing detector}} & \textbf{ACC} & \textbf{F1} & \textbf{Precision} & \textbf{Recall} & \textbf{Time} \\
\midrule  
\multirow{4}{*}{\rotatebox{90}{Traditional}} & HTMLPhish & 52.86 & 42.18 & 64.01 & 53.44 &  $\ll 0.1$ s \\
& Web2Vec & 50.55 & 38.64 & 55.29 & 51.02 &  $\approx 0.1$ s \\
& HTML-GNN & 74.50 & 74.19 & 76.11 & 74.74 &  $< 0.1$ s \\
& PhishDet & 72.11 & 71.98 & 72.38 & 72.06 &  $\ll 0.1$ s \\
\midrule    
\multirow{3}{*}{\rot{GNN}} 
& DAG & 77.93 & 77.84 & 78.46 & 77.90 & $\ll 0.1$ s \\
& GNN$^+$ & 78.73 & 78.47 & 79.40 & 78.69 &  $\ll 0.1$ s \\
& PIGVAE & 70.05 & 69.69 & 71.04 & 70.05 &  $\approx 0.4$ s \\
\midrule  
\multirow{6}{*}{\rot{Ref-based}}
& \multirow{2}{*}{Phishpedia} & 70.17 & 67.96 & 77.88 & 70.17 &  $\approx 0.3$ s \\
& & 66.70 & 62.69 & 79.28 & 66.70 & $\approx 0.4$ s \\\cmidrule(lr){2-7}
& \multirow{2}{*}{\textcolor{red}{PhishIntention}} & \textcolor{red}{86.63} & \textcolor{red}{86.42} & \textcolor{red}{89.12} & \textcolor{red}{86.63} & \textcolor{red}{\underline{$\mathbf{\approx}$ \textbf{5 s}}} \\
& & 84.26 & 83.87 & 87.97 & 84.26 & $\approx 6$ s \\\cmidrule(lr){2-7}
& \textcolor{red}{PhishLLM} & \textcolor{red}{87.50} & \textcolor{red}{87.31} & \textcolor{red}{90.00} & \textcolor{red}{87.50} & \textcolor{red}{\underline{$\mathbf{\approx}$ \textbf{3 s}}} \\ 
\midrule   
& \textcolor{blue}{SpecularNet} & \textcolor{blue}{86.06} & \textcolor{blue}{86.05} & \textcolor{blue}{86.16} & \textcolor{blue}{86.06} & \textcolor{blue}{\underline{\textbf{10 ms}}} \\
\bottomrule
\end{tabular}
\end{adjustbox}
\label{tab:perf_2}
\renewcommand{\arraystretch}{1.0}
\end{table}

\begin{table}[t!]
\scriptsize
\centering
\caption{Results of ablation study on \emph{knowphish-dataset}.}
\vspace{-1.5em}
\resizebox{\linewidth}{!}{
\renewcommand{\arraystretch}{0.85}
\begin{tabular}{p{2.8cm}|p{0.6cm}<{\centering}p{0.6cm}<{\centering}p{0.8cm}<{\centering}p{0.6cm}<{\centering}p{0.6cm}}
\toprule
\textbf{Description} & \textbf{ACC} & \textbf{F1} & \textbf{Precision} & \textbf{Recall} \\
\midrule  
Original SpecularNet & 93.92 & 93.92 & 94.02 & 93.91  \\
\midrule
w/o classification loss & 92.21 & 92.51 & 92.21 & 92.20  \\
w/o reconstruction error loss & 92.61 & 92.61 & 92.73 & 92.60 \\
w/o graph decoder & 91.32 & 91.36 & 91.32 & 91.32 \\
w/o graph AE (only GCN) & 82.14 & 82.16 & 82.14 & 82.13  \\  
w/o base GNN (GCN$\times 3$) & 92.75 & 92.93 & 92.75 & 92.74 \\
w/o domain names & 90.14 & 89.99 & 90.64 & 90.08 \\ 
\bottomrule
\end{tabular}}
\label{tab:ablation}
\renewcommand{\arraystretch}{1.0}
\end{table}

\begin{table*}[t!]
\centering
\footnotesize
\setlength{\tabcolsep}{6pt} % adjust if needed
\caption{Field and robustness evaluation of SpecularNet under realistic deployment conditions.}
\vspace{-1.0em}
\begin{tabular}{|
  >{\centering\arraybackslash}p{0.32\textwidth} |
  >{\centering\arraybackslash}p{0.3\textwidth} |
  >{\centering\arraybackslash}p{0.3\textwidth} |
}
\toprule
% ---------- Col 1 ----------
\textbf{\normalsize{Table 5.1: Field study 1.}} &
\textbf{\normalsize{Table 5.2: Field study 2.}} &
\textbf{\normalsize{Table 5.3: Adversarial attacks.}} \\[-0.7em]
% ---------- Inner tables row ----------
\begin{tabular}[t]{@{}lcc@{}}
\midrule
\textbf{Detector} & \multicolumn{2}{c}{\textbf{\# revealed}} \\
\midrule
DynaPhish/PhishLLM & 1388 & 1029 \\[0.4em]
SpecularNet & 1350 (97.26\%) & 981 (95.34\%) \\
\end{tabular}
&
\begin{tabular}[t]{@{}lcccc@{}}
\midrule
 & \textbf{ACC} & \textbf{F1} & \textbf{Prec.} & \textbf{Rec.} \\
\midrule
SpecularNet & 81.25 & 81.22 & 81.47 & 81.25 \\
\end{tabular}
&
\begin{tabular}[t]{@{}lcc@{}}
\midrule
\textbf{Dataset} & Zenodo & $\delta$phish \\
\midrule
No attack & 94.90 & 90.82 \\
Simple attack & 94.90 & 83.67 \\
Advanced attack & 94.39 & 73.47 \\
\end{tabular}
\\
\bottomrule
\end{tabular}
\label{tab:field_adver_triplet}
\end{table*}

\subsection{Detection Performance \& Benchmarking}

Tables~\ref{tab:perf_1} and~\ref{tab:perf_2} report the detection performance and inference time per webpage on the \emph{knowphish-dataset} and \emph{phishllm-dataset}, respectively, which represent temporally and distributionally shifted evaluation settings relative to the 2021 training data. Across both test sets, SpecularNet consistently outperforms traditional DL baselines and generic GNNs, while achieving competitive performance with state of the art reference-based detectors. On \emph{knowphish-dataset}, SpecularNet attains an F1 score of 93.92\%, ranking second overall and trailing the best reference-based method by less than 1.5 percentage points. On \emph{phishllm-dataset}, which introduces additional distribution shift, SpecularNet maintains strong performance with an F1 score of 86.05\%, ranking among the top three detectors.

Traditional models such as HTMLPhish and Web2Vec exhibit substantial performance degradation under temporal shift, despite strong results under conventional random splits, indicating limited generalization. Generic GNNs improve upon these baselines but remain consistently inferior to SpecularNet, highlighting the benefit of task-specific modeling of DOM tree structure.

While reference-based approaches still achieve the highest absolute accuracy, SpecularNet delivers comparable results without relying on external knowledge bases, multimodal pipelines, or cloud services. The remaining performance gap on \emph{phishllm-dataset} is partly explained by the absence of domain name information, as confirmed by our ablation study and prior work~\cite{liu2024less}.

Crucially, SpecularNet offers orders-of-magnitude faster inference, requiring approximately 20 ms per webpage, including DOM parsing from raw HTML (true end-to-end latency), compared to 2–7 seconds for leading reference-based systems. Inference time scales primarily with DOM size and CPU capability: for typical webpages (median $\sim 1,000$ nodes), latency is negligible, and even for the 90th percentile ($\sim 5,000$ nodes), it remains below 0.1 seconds on commodity CPUs. Extremely large pages incur higher latency but are rare in practice. In contrast, referenc--based detectors incur largely device-independent latency dominated by external dependencies.

\subsection{Ablation Study}

We conduct a component-wise ablation study by removing or modifying one module at a time and evaluating performance on \emph{knowphish-dataset} (\emph{phishllm-dataset} is excluded due to the absence of domain name information), as summarized in Table~\ref{tab:ablation}. Specifically, we examine: (i) removing the classification loss while retaining the reconstruction loss, and vice versa; (ii) discarding the decoder and using the root node embedding directly for classification; (iii) removing the autoencoding mechanism while preserving the GCN layers; (iv) excluding the base GNN and directly applying top-$k$ pooling to the tree; and (v) omitting the domain node.

Across all settings, performance consistently degrades when any component is removed, confirming the contribution of each design choice. The most pronounced decline occurs when the autoencoder is entirely omitted, underscoring the central role of hierarchical graph autoencoding in SpecularNet’s effectiveness.

\subsection{Field Studies -- Detection in the Wild}

To evaluate the effectiveness of SpecularNet against phishing attacks in the wild beyond benchmark datasets, we follow the methodology of DynaPhish~\cite{liu2023knowledge} and PhishLLM~\cite{liu2024less}, and measure how many emerging real-world phishing webpages identified by reference based detectors are also detected by SpecularNet. Both papers released field-study datasets consisting of manually validated phishing websites associated with newly issued or updated TLS certificates, collected from CertStream~\cite{certstream}. Table~\ref{tab:field_adver_triplet}.1 reports the results. SpecularNet successfully detects up to 97\% of phishing webpages uncovered by reference-based approaches, achieving comparable effectiveness under real-world conditions.

To further assess robustness in a fully open setting, we additionally construct a recent, independent dataset comprising 6,000 newly observed websites, evenly split between benign domains sampled from the latest Cloudflare domain rankings~\cite{cloudflare_radar_domains} and active phishing websites (as of 2026-01-19) obtained from the open feed of \textit{Phishing.Database}~\cite{phishing_database_github}, an official data provider for VirusTotal~\cite{virustotal_upload}. Table~\ref{tab:field_adver_triplet}.2 reports the classification results. Despite being trained on data collected nearly five years earlier, SpecularNet maintains strong performance, with an expected degradation of approximately 10\%, thereby demonstrating robust temporal generalization and resilience in real-world deployments.

\subsection{Robustness Against Adversarial Attacks}

Adversarial manipulation of phishing webpages, often implemented through HTML obfuscation or structural perturbations, is a common strategy used by attackers to evade ML-based phishing detectors. Such attacks typically modify specific HTML patterns or structural cues exploited by learning-based models, while preserving the visual appearance and functionality of the phishing page.

Our adversarial evaluation focuses on HTML-level evasion, which aligns with both the threat model of SpecularNet and realistic attacker behavior. In particular, we adopt SpacePhish~\cite{apruzzese2022spacephish}, one of the most established frameworks for generating adversarial phishing webpages through principled HTML manipulations. The perturbations introduced by SpacePhish emulate realistic attacker strategies, such as structural noise injection and tag-level modifications, without relying on visual or semantic changes. Reference-based approaches are excluded from this evaluation due to their reliance on external references and visual cues, which makes them insensitive to HTML-level perturbations. We therefore evaluate SpecularNet on the SpacePhish datasets ($\delta$phish and Zenodo), which explicitly target HTML-based adversarial manipulation.

The results in Table~\ref{tab:field_adver_triplet}.3 show that while adversarial perturbations on $\delta$phish lead to a measurable performance drop, SpecularNet remains robust overall. Notably, prior work reports that SpacePhish-based attacks can degrade detection accuracy by up to 50\%~\cite{apruzzese2022spacephish}, whereas SpecularNet retains a substantially higher accuracy. This resilience can be attributed to SpecularNet's reliance on DOM-level structural invariants, which are inherently more difficult to manipulate without compromising the visual fidelity or usability of phishing webpages.

\section{Conclusion}
We presented SpecularNet, a lightweight phishing detection framework that addresses the limitations of conventional learning-based methods while providing a practical alternative to highly complex reference-based systems. By relying exclusively on a webpage's domain name and HTML structure, SpecularNet introduces a novel hierarchical graph autoencoding model to effectively exploit the inherent properties of the DOM tree. Extensive evaluation on three validated datasets demonstrates that SpecularNet achieves detection performance comparable to state-of-the-art reference-based approaches, while requiring only minimal inference time and no external dependencies. Beyond benchmark datasets, SpecularNet exhibits strong robustness in field studies on real-world phishing websites and maintains resilience against realistic HTML-level adversarial manipulations, further supporting its suitability for deployment under evolving threat conditions. These results highlight the potential of SpecularNet as an effective, scalable, and fully reference-free solution for real-world phishing detection.

\begin{acks}
This work has been supported by the FWF Austrian Science Fund, Project reference I-6653, as part of the EU CHIST-ERA-2022-SPiDDS-02 project \emph{GRAPHS4SEC}. The work has been also partially supported by Cisco Systems Inc., the European Union under the Italian National Recovery and Resilience Plan (NRRP) of NextGenerationEU, partnership on “Telecommunications of the Future” (PE00000001 - program “RESTART”, Focused Project R4R), and the SmartData@PoliTO center on Big Data and Data Science.
\end{acks}

\bibliographystyle{plain}


% \newpage
\appendix

\section{Dataset detail}
We use publicly available datasets to ensure a fair, direct, and reliable comparison.

\subsection{Training set -- \emph{phishpedia-dataset}}
We use the \emph{phishpedia-dataset} as the sole training set not because it contains the largest number of samples, but because it is the oldest, allowing us to evaluate generalizability over time.

\subsection{Test set 1 -- \emph{knowphish-dataset}}
Thirteen samples from the \emph{knowphish-dataset} are either corrupted or inherently flawed, so we have to remove them. This is considered a trivial issue, as their impact on overall performance is negligible --- even if none of them were correctly classified.

The results of reference-based approaches on test set 1 are directly taken from Table 2\footnote{``KnowPhish: Large Language Models Meet Multimodal Knowledge Graphs for Enhancing Reference-Based Phishing Detection" } and Table 1\footnote{``PhishAgent: A Robust Multimodal Agent for Phishing Webpage Detection"} in their respective original papers, as both works employed the ``same" dataset. However, the authors of the first work evaluated two separate datasets, in which one contains two-fifths of test set 1 and the other comprises benign samples from the remaining three-fifths of test set 1 alongside phishing samples sourced elsewhere. Therefore, we deduce the final performance through the weighted sum of both subsets\footnote{For example, $ACC_{final}=ACC_{TR-OP}\times 0.4+ACC_{TR-AP}\times 0.6$, where TR-OP (as named in the original paper) includes 4,000 samples identical to our test set 1, and TR-AP (dataset name) contains the remaining 3,000 benign samples in our test set 1 along with 3,000 externally sourced phishing samples, which are unavailable to us. Even with this calibration, the performance reported in our paper does not deviate substantially from the best results presented in the original publications, thereby affirming the competitiveness of SpecularNet.} for a fair comparison with GEPAgent, ChatPhish, and PhishAgent. The results of other reference-based models are left unaltered, as the second work uses the intact test set 1.

\subsection{Test set 2-- \emph{phishllm-dataset}}
The results of reference-based approaches on test set 2 are derived based on Table 3\footnote{``Less Defined Knowledge and More True Alarms: Reference-based Phishing Detection without a Pre-defined Reference List"} in the original paper. The authors only showcases the precision and recall of phishing webpages, but as the quantities of both classes are provided, we can safely infer the overall evaluation metrics reported in our paper. Notably, the released dataset contains slightly more than 6,075 samples per class, though the reason for this discrepancy is unclear. To ensure a fair comparison, we randomly sample the required number of instances from each class. Importantly, even when including the randomly discarded samples, our model's performance remains consistent. Furthermore, since the phishing samples in test set 2 are generated using phishing kits and lack associated domain names, we retrain our model with all domain-related components removed.

\subsection{Dataset for field study}
For the first field study, both datasets are provided by their respective original papers and consist of real-world phishing webpages identified through the authors' proposed reference-based approaches. The phishing websites have been validated by multiple domain experts. The number of provided webpages slightly differs from what is reported in the original papers (for reasons unknown to us), so we can only rely on the available data for performance evaluation.

\subsection{Dataset for adversarial attack}
The dataset used in our work is directly sourced from the samples (without attack / with simple attack / with advanced attack) provided in the original paper’s GitHub repository. 

\section{Hyperparameter range}
Table~\ref{tab:hyper_range} lists the range of each hyperparameter for model tuning.

\begin{table}[h]
\footnotesize
\centering
\renewcommand{\arraystretch}{1}
\caption{Details about hyper-parameters.}
\scalebox{0.97}{
\begin{tabular}{p{3.6cm}|p{3.1cm}<{\centering}}
\toprule
\textbf{Parameter} & \textbf{Range} \\
\midrule  
Feature dimension, $F$ & [16, 32, 64] \\
Num layers of GCN & 1-6 \\
Size of hidden channel in GCN & [16, 32, 64, 128] \\
Num layers of LSTM & 1-3 \\
Size of hidden channel in LSTM & [16, 32, 64] \\
Ratio for topK pooling & [5\%, 10\%, 15\%, 20\%, 25\%, 30\%] \\
Num neurons of linear layer in AE & [16, 32, 64] \\
Num neurons of classification MLP & [(16, 1), (32, 1), (16, 8, 1), (32, 16, 1)] \\
Activation function & [ReLU, LeakyReLU] \\
Optimizer & [SGD, Adam] \\
Batch size & [8, 16, 32, 64] \\
Initial learning rate & [$10^{-2}$, $10^{-3}$, $10^{-4}$] \\
\bottomrule
\end{tabular}}
\label{tab:hyper_range}
\end{table}

\section{Training time}
Indeed, the training-free nature of state of the art reference-based approaches constitutes a unique advantage in today’s cyber defense landscape, wherein the rapid evolution of phishing activities may necessitate continuous model retraining. Fortunately, SpecularNet requires merely less than one hour to converge, let alone the non-necessity for GPU acceleration.

\section{About ensemble classification}
\begin{figure}[H]
\centering
\includegraphics[scale=0.6]{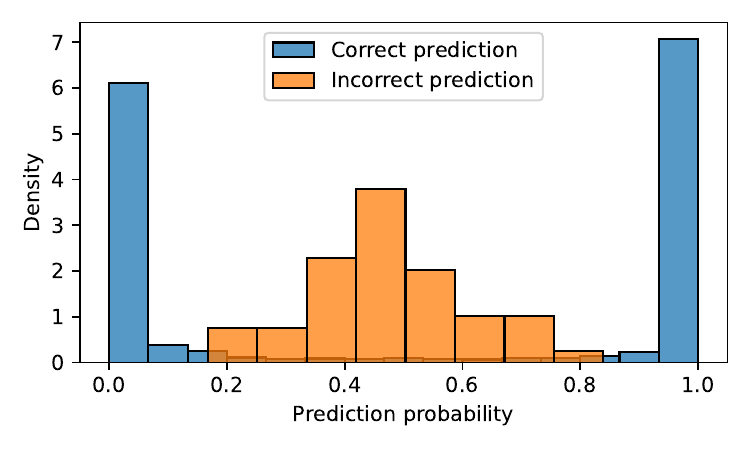}
\caption{Prediction probability for correct and incorrect predictions (blue: error-mapping and threshold-based classifications are both correct \& orange: error-mapping classifications are wrong but threshold-based ones are correct).}
\label{fig:prob}
\end{figure}

Figure~\ref{fig:prob} shows the prediction probabilities of classifications based on error mapping, conditioned on the threshold-based classifications being correct. When both classification methods are correct, the model demonstrates high confidence in its predictions. However, when the error-mapping classification errs, the model’s confidence decreases; simultaneously, the threshold-based classification remains correct, compensating for the error and thereby improving overall performance. This also indicates that the two classification methods can discern varying degrees of phishing/benign patterns, thereby complementing each other.

\section{Regarding the label convention}
We follow the common labeling convention where class 0 ($y=0$) denotes benign/positive samples, and class 1 ($y=1$) denotes malicious/phishing/anomalous/negative samples. However, Equation 6 adopts the inverse convention, assigning 0 to phishing and 1 to benign, which is only used during model training.

\end{document}